\documentclass{aa}  
\usepackage{graphicx,ulem}
\usepackage{txfonts}
\usepackage{natbib}
\bibpunct{(}{)}{;}{a}{}{,}
\begin{document}
   \title{High spatial resolution mid-infrared observations of the low-mass young star TW~Hya           
  \thanks{Based on observations with the Very Large Telescope
   Interferometer (VLTI, proposal 075.C-0014).}}

   \subtitle{}

   \author{Th. Ratzka\inst{1,2}
           \and
           Ch. Leinert\inst{2}
           \and
           Th. Henning\inst{2}
          \and
           J. Bouwman\inst{2}
          \and
           C. P. Dullemond\inst{2}
          \and
           W. Jaffe\inst{3}
          }

   \offprints{tratzka@aip.de}

   \institute{
	      Astrophysical Institute of Potsdam,
	      An der Sternwarte 16, D-14482 Potsdam,
	      Germany
              \and
              Max-Planck-Institute for Astronomy,
              K\"onigstuhl 17, D-69117 Heidelberg,
              Germany
	      \and
	      Sterrewacht Leiden, P.O. Box 9513, 
              NL-2300 RA Leiden, The Netherlands
             }

   \date{Received Feb 26, 2007; accepted Apr 25, 2007}

 
  \abstract
   {}
   {We want to improve knowledge of the structure of the inner few AU of the circumstellar disk around the nearby T~Tauri star TW~Hya. Earlier studies have suggested the existence of a large inner hole, possibly caused by interactions with a growing protoplanet.}
   {We used interferometric observations in the N-band obtained with the MIDI instrument on the Very Large Telescope Interferometer, together with 10\,$\mu$m spectra recorded by the infrared satellite Spitzer. The fact that we were able to determine N-band correlated fluxes and visibilities for this comparatively faint source shows that mid-infrared interferometry can be applied to a large number of low-mass young stellar objects.}
   {The mid-infrared spectra obtained with Spitzer reveal emission lines from H\,I (6-5), H\,I (7-6), and [Ne\,II] and show that over 90\% of the dust we see in this wavelength regime is amorphous. According to the correlated flux measured with MIDI, most of the crystalline material is in the inner, unresolved part of the disk, about 1\,AU in radius. The visibilities exclude the existence of a very large ($3-4$\,AU radius) inner hole in the circumstellar disk of TW Hya, which was required in earlier models. We propose instead a geometry of the inner disk where an inner hole still exists, but at a much reduced radius, with the transition from zero to full disk height between 0.5 and 0.8\,AU, and with an optically thin distribution of dust inside. Such a model can comply with SED and mid-infrared visibilities, as well as with visibility and extended emission observed in the near-infrared at 2\,$\mu$m. If a massive planet was the reason for this inner hole, as has been speculated, its orbit would have to be closer to the star than 0.3\,AU. Alternatively, we may be witnessing the end of the accretion phase and an early phase of an inward-out dispersal of the circumstellar disk.}
   {}

   \keywords{Stars: individual: TW Hya --
             Stars: circumstellar matter --
   	     Stars: pre-main sequence --
             Techniques: interferometric --
   	     Infrared: stars
             }

   \maketitle
%

\section{Introduction}

The classical T~Tauri star TW~Hya is the most prominent member of the nearby TW~Hya association. With its distance of $51\pm 4$\,pc \citep{mamajek05} and its estimated age of $\sim 10$\,Myr \citep{hoff98,webb99} TW~Hya is a Rosetta stone for investigating the transition phase from disks to planets. The object shows clear signs of ongoing accretion, but the rate -- e.g.~\cite{muzerolle00} derived $\sim 4\cdot10^{-10}\,\rm{M}_{\odot}\,\rm{yr}^{-1}$ -- is much lower than the typical value of $10^{-8}\,\rm{M}_{\odot}\,\rm{yr}^{-1}$ found for 1\,Myr old classical T~Tauri stars in Taurus \citep{calvet00}. Images taken at various wavelengths from the optical to the millimetre regime reveal the presence of a nearly face-on optically thick dust disk; see \cite{weinberger02} and references therein. \cite{apai04} show the presence of polarised disk emission in the K-band as close as 0\farcs1 ($\sim 5$\,AU) to the central star by using the polarimetric differential imaging mode of the adaptive optics system NACO at the VLT. 

A detailed analysis of the spectral energy distribution (SED) led \cite{calvet02} to conclude that the inner disk region within 4\,AU of the central star is almost devoid of dust, with only an optically thin layer of small (sub)micron-sized dust grains remaining. Such a region can be produced by some mechanisms of dust depletion like photoevaporation, inner planet(s), or significant grain coagulation. Indeed, millimetre observations of the outer disk point to a significant fraction of even centimetre-sized dust grains \citep{wilner05}. However, even with HST imaging and sophisticated AO techniques, it was not possible to characterise the innermost regions of the disk. \cite{rettig04} were able to detect rovibrationally excited CO from the inner disk but with no evidence of very hot gas. They conclude that TW Hya has cleared its inner disk out to a radial distance of $\sim$ 0.5\,AU and is approaching the end of its accretion phase. With Keck interferometry measurements, \cite{eisner06} could spatially resolve the dust continuum emission at a wavelength of 2\,$\mu$m. They concluded that the inner disk consists of optically thin, submicron-sized dust extending to within 0.06\,AU where it may be magnetospherically truncated.

In this paper we present the first interferometric measurements of the TW~Hya disk in the thermal infrared. The observations outlined in Sect.~\ref{obs} have been performed with the `MID-infrared Interferometric instrument' (MIDI) attached to the VLTI \citep{leinert03b, leinert03a}. The data analysis is explained in Sect.~\ref{dr}. In Sect.~\ref{res}, the reduction of spectra obtained with the `Spitzer Space Telescope' is described. These spectra are used to derive the global dust composition of the mid-infrared emitting region (Sect.~\ref{composition}) that serves as input for our modified Chiang \& Goldreich model introduced in Sect.~\ref{model}. By fitting the SED and the visibility (Sect.~\ref{sed_visibility}), the geometrical properties of the circumstellar disk can be analysed. The results and their implications for our picture of this young stellar object are dicussed in Sect.~\ref{discussion}. Finally, in Sect.~\ref{conclusion} we summarise our findings. 

%

\section{Interferometric observations\label{obs}}

\begin{table*}
\caption{Journal of observations.}
\label{table1}
\centering
\begin{tabular}{cccccccccc}
\hline\hline
\noalign{\smallskip}
Date of     & Universal     & Object    & $F_{10\,\mu{\rm m}}$ & \multicolumn{2}{c}{Projected Baseline} &  Airmass & Acquisition & Interferometric & Photometric\\
Observation & Time          &           & [Jy]           & \ \ [m]\ \ & [deg]                     &          & Frames      & Frames	   & Frames\\
\noalign{\smallskip}
\hline
\noalign{\smallskip}
28-05-2005  & 22:22 - 23:34 & HD 102461 &   12.6 & 50.4 & 24.4 & 1.20 & 1000$\times$4\,ms & \ \ 8000$\times$18\,ms  & 2$\times$4000$\times$18\,ms \\
29-05-2005  & 00:20 - 00:44 & HD 102839 &\ \  -  & 42.4 & 39.9 & 1.44 & 1000$\times$4\,ms & \ \ 8000$\times$18\,ms  & 2$\times$4000$\times$18\,ms \\
29-05-2005  & 01:10 - 01:41 &         HD 95272   &\ \ 9.5 & 55.5 & 35.7 & 1.20 & 4000$\times$4\,ms & \ \ 8000$\times$18\,ms  & 2$\times$4000$\times$18\,ms \\
29-05-2005  & 01:41 - 02:19 &         TW Hya     & \ \  - & 49.8 & 41.0 & 1.30 & 6000$\times$4\,ms &    12000$\times$22\,ms  & 2$\times$6000$\times$22\,ms \\
29-05-2005  & 02:19 - 02:46 &         TW Hya     &\ \   - & 47.9 & 42.4 & 1.42 & 4000$\times$4\,ms &    12000$\times$22\,ms  & 2$\times$6000$\times$22\,ms \\
29-05-2005  & 02:46 - 03:12 &         HD 95272   &\ \ 9.5 & 51.6 & 37.4 & 1.72 & 1000$\times$4\,ms & \ \ 8000$\times$22\,ms  & 2$\times$4000$\times$22\,ms \\
29-05-2005  & 03:41 - 04:07 & HD 139127 &   16.3 & 54.3 & 27.5 & 1.05 & 1000$\times$4\,ms & \ \ 8000$\times$18\,ms  & 2$\times$4000$\times$18\,ms \\
29-05-2005  & 05:17 - 05:42 & HD 133774 &   11.5 & 55.9 & 35.3 & 1.20 & 1000$\times$4\,ms & \ \ 8000$\times$18\,ms  & 2$\times$4000$\times$18\,ms \\
\noalign{\smallskip}
\hline
\noalign{\smallskip}
\multicolumn{10}{c}{\parbox{17.6cm}{Notes: All observations used the baseline UT1-UT2. The length and the position angle of the projected baseline refer to the time of the fringe tracking and the airmass to the time of the photometry. The 10\,$\mu$m-fluxes of the spectrophotometric calibrators have been taken from \cite{cohen99}.}}\\
\noalign{\smallskip}
\hline
\end{tabular}
\end{table*}

For a spatially and spectrally resolved study of TW~Hya and its dusty environment in the N-band, we observed this source with MIDI. This instrument allows an interferometric combination of the light coming from a pair of telescopes \citep{leinert03b, leinert03a, morel04, ratzka05, ratzka06}.

The two consecutive measurements analysed and discussed in this paper were obtained with the baseline formed by the two Unit Telescopes Antu (UT1) and Kueyen (UT2) on 28 May 2005 within the scope of the MIDI `Guaranteed Time Observations' (GTO), i.e.\ telescope time dedicated to the MIDI consortium. The actual baseline length due to the projection on the sky was about 50\,m and thus provided a spatial resolution or, more precisely, fringe spacing of about 0\farcs04 at 10\,$\mu$m. This corresponds to 2\,AU at the distance of TW~Hya. A journal of observations is given in Table~\ref{table1}.

\subsection{Observing sequence\label{os}}

After coarse acquisition by the telescopes, chopped images of the source were taken. These images with an angular diameter of about 2\arcsec allow adjustment of the position of the object to a predetermined pixel on the detector, a $320 \times 240$ pixel Si:As Impurity Band Conduction (IBC) array, in order to maximise the overlap of both telescope beams for the subsequent interferometric measurements. 

Then the beam combiner, which produces two interferometric outputs of opposite sign, was put into the optical train. An important feature of the measuring process with MIDI is the possibility of using a dispersive element. We used a prism with a low spectral resolution of $\Delta\lambda/\lambda\sim30$, together with a 0\farcs52 wide slit. This spectral resolution not only allowed us to obtain {\it dispersed fringes}, i.e.~spectrally resolved interferograms, but also increased the coherence length to about $\pm$~300~$\mu$m, corresponding to an interferometric field of view of 1\farcs2 for the used projected baseline length. Imperfect centring on the white-light fringe within a few $\lambda$ thus had no adverse effect on the determination of the visibility. 

After finding the location of zero optical path difference (OPD) by scanning a few millimetres around the expected point of path length equalisation, an interferometric measurement with self-fringe-tracking was started. In this mode, the piezo-mounted mirrors within MIDI are typically used to scan a range in OPD of eight wavelengths ($\lambda\sim 10$\,$\mu$m) in steps of $2$\,$\mu$m. At each step, with the corresponding, fixed OPD, an exposure is taken that gives the instantaneous value of the interferometric signal from $8$ to $13$\,$\mu$m. After each scan of 80\,$\mu$m, the position of the fringe packet in the scan is determined and the VLTI delay lines adjusted in order to recentre the fringe packet for the next scan. The scans can either be centred on the white-light fringe or at a fixed offset from zero OPD. The latter is advantageous for the coherent data evaluation (Sect.~\ref{dr}), which measures the spectral modulation of the fringe at all wavelengths simultaneously and thus is subject to confusion with background emission at zero OPD. The output after a scan is the spectrally resolved temporal fringe pattern, which gives the fringe amplitude or correlated flux. The scans are repeated in a saw-tooth manner typically $100-200$ times to increase the statistical accuracy. No chopping is necessary for these interferometric measurements, because the two resulting output signals can be subtracted from each other, which gives an efficient first-order background correction.
 
Finally, to derive the (wavelength-dependent) visibilities, spectra were recorded on the same pixels of the detector as the fringe signal by blocking first the light from one, then from the other telescope while chopping the secondary mirror of the UTs to remove the thermal background radiation. By definition the visibility is obtained as the ratio of correlated flux from the interferometric measurement and total flux from the photometry (Sect.~\ref{dr}). This gives the raw visibility, still suffering from atmospheric and instrumental correlation losses.

The observing sequence for an interferometric measurement with MIDI on the VLTI thus produces a diffraction-limited image of the object within a certain wavelength range defined by a filter, a spectrum from 8 to 13 $\mu$m, and the correlated fluxes, i.e.~the visibilities, over the same wavelength range.

\subsection{Calibration}

Since the strength of the fringe amplitude can be reduced by various instrumental and atmospheric effects, calibrators of known intrinsic visibility have to be observed to determine the instrumental visibility, i.e.~the raw visibility that would be observed on a perfect point source. The instrumental visibility can be understood as the wavelength-dependent transfer function of the instrument. Due to the additional correlation losses caused by atmospheric turbulence, the calibration measurements should be timely and spatially as close as possible to those of the scientific target. 

\begin{figure}
\centering
\includegraphics[height=9cm,angle=90]{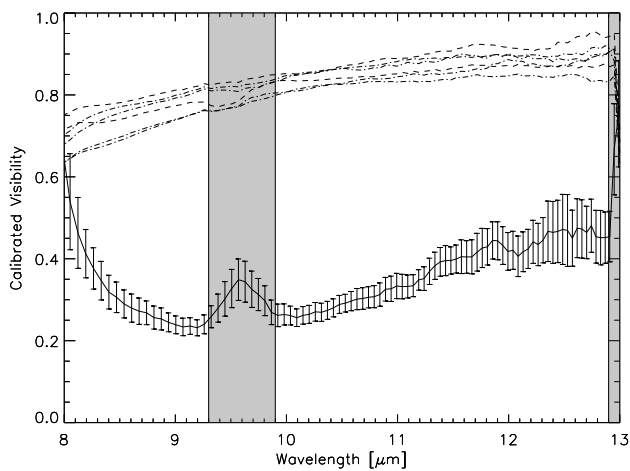}
\caption{Calibrated visibility of TW~Hya. The solid curve with error bars shows the average of the two independent measurements of TW~Hya and the associated calibrator HD~95272. The instrumental visibilities derived from the individual observations of HD~95272 (dashed) and the other calibrators (dashed-dotted) are also shown. A smoothing of 5 pixels in wavelength direction has been applied to all measurements. The regions between 9.3\,$\mu$m and 9.9\,$\mu$m and above 12.9\,$\mu$m are affected by the atmosphere.} 
\label{fig1}
\end{figure}

Therefore, we obtained measurements of the calibrator HD~95272 immediately before and after the measurements of TW~Hya by performing the same observation sequence that is described in Sect.~\ref{os}. The resulting raw visibilities were then corrected for the finite size of the calibrator by assuming a uniform disk with the known stellar diameter. Other calibrators observed in the same mode on the same baseline during the same night (see Table~\ref{table1}) were used only to check the stability of the instrumental visibility (Fig.~\ref{fig1}). This uncertainty also contributes to the error bars of the final visibility of TW~Hya, but it is smaller than the uncertainties in the interferometric measurements of the scientific target.

The interferometric calibrators were taken from the `MIDI Calibrator Catalogue'$^1$ of 509 stars that are at least 5\,Jy bright at $10$\,$\mu$m and selected for the absence of circumstellar emission, disturbing companions, or strong variability. A subsample$^2$ of these stars can also be used for absolute flux calibration. With the exception of HD~102839, all calibrators in Table~\ref{table1} are such spectrophotometric standards. However, we restricted ourselves to measurements of HD~95272 for determining the spectrum.
\footnotetext[1]{http://www.eso.org/$\sim$arichich/download/vlticalibs-ws/}
\footnotetext[2]{http://www.eso.org/instruments/midi/tools/spectrophot\_std.html}
%

\section{MIDI Data reduction and results\label{dr}}

\begin{figure}
\centering
\includegraphics[height=9cm,angle=90]{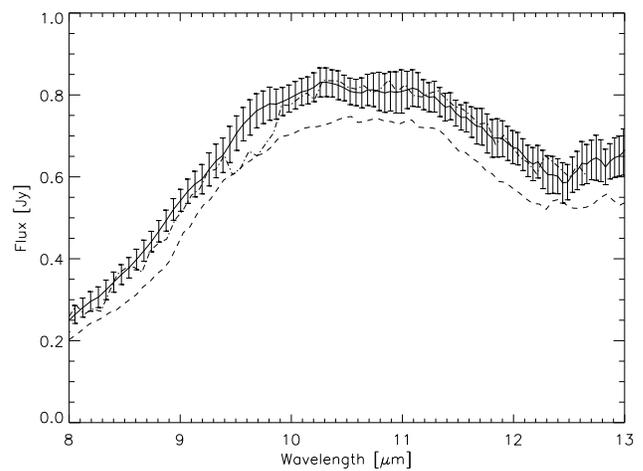}
\caption{Spectrum of TW~Hya as determined by MIDI (solid), Spitzer (dashed), and by \cite{weinberger02} with Keck (dashed-dotted). The errors of the Spitzer data are negligible and those of the Keck spectrum are in general smaller by a factor of about two when compared to those of the MIDI spectrum. They have not been plotted. A smoothing of 5 pixels in wavelength direction has been applied to the MIDI results.}  
\label{fig2}
\end{figure}

\begin{figure}
\centering
\includegraphics[height=9cm,angle=90]{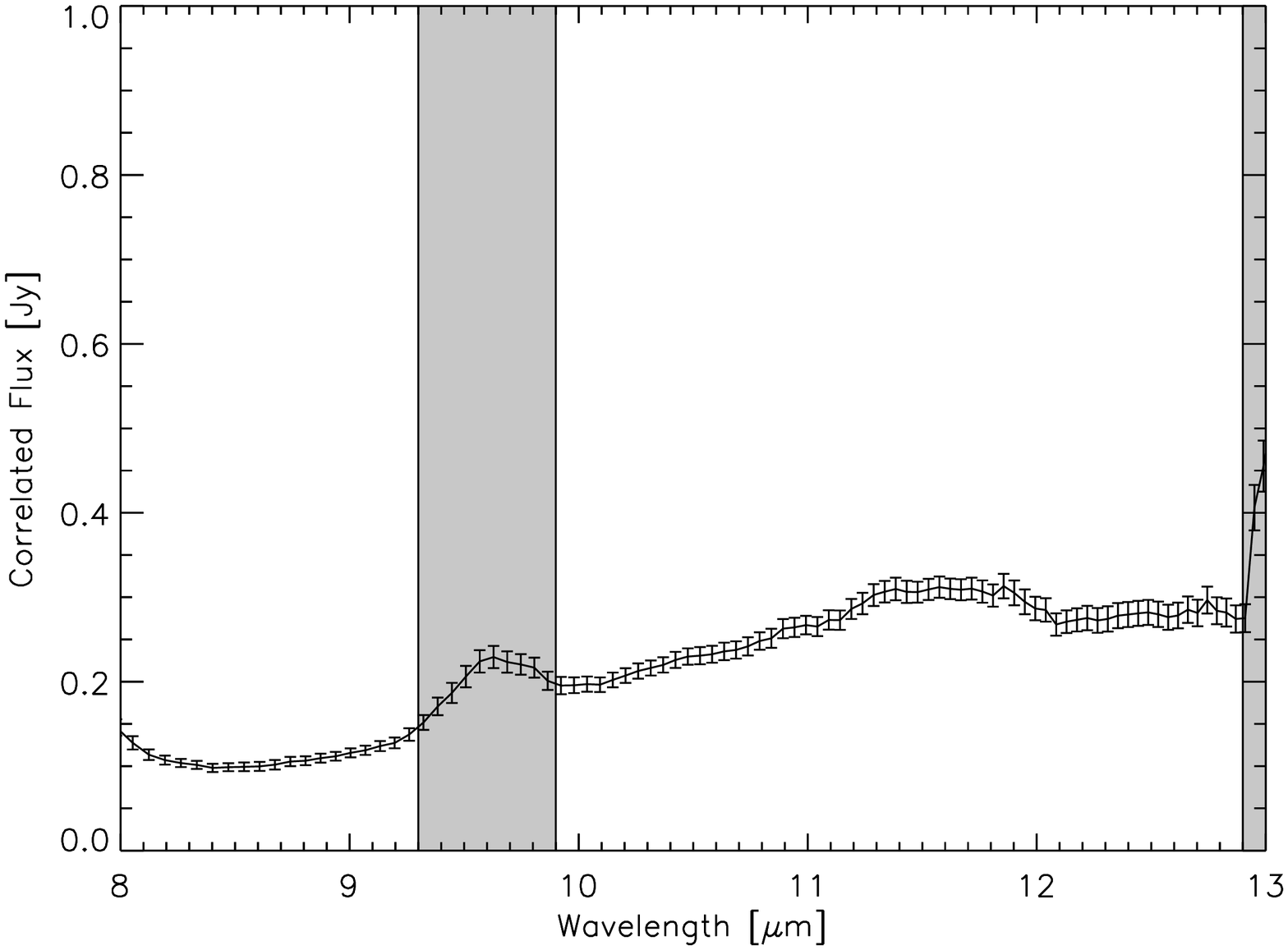}
\caption{Directly calibrated correlated flux. A smoothing of 5 pixels in wavelength direction has been applied. The regions between 9.3\,$\mu$m and 9.9\,$\mu$m and above 12.9\,$\mu$m are affected by the atmosphere.}  
\label{fig3}
\end{figure}

\subsection{Visibilities\label{drint}}

For the data reduction of the interferometric measurements, the software package {\it MIA+EWS}$^{3,4}$ written in C and IDL was used. The correlated flux $F^{\rm corr}\left(\lambda\right)$ and the total flux $F^{\rm total}\left(\lambda\right)$ as well as their ratio, i.e. the wavelength-dependent visibility
\begin{equation}
V_{raw}(\lambda) = {F^{\rm corr}\left(\lambda\right)\over F^{\rm total}\left(\lambda\right)}{\rm ,} \label{eq1}
\end{equation}
can be determined with these programs. The first branch of the data reduction software called {\it MIA} is based on power spectrum analysis, i.e.~the determination of the spatial modulation of the fringe in distinct wavelength bands. The necessary data reduction steps are described in a tutorial on the cited web pages, as well as in \cite{leinert04} and \cite{ratzka05}. The second branch, called {\it EWS}, uses a shift-and-add algorithm in the complex plane, i.e.~averages over suitably modified individual exposures, to obtain the complex visibility $A(\lambda)\cdot e^{i\phi(\lambda)}$. Thus this is a coherent and linear algorithm. \cite{jaffe04} and the documentation on the web page give further insight into the program.

\footnotetext[3]{http://www.strw.leidenuniv.nl/$\sim$koehler/MIA+EWS-Manual}
\footnotetext[4]{http://www.mpia-hd.mpg.de/MIDISOFT}

In contrast to MIA, which takes advantage of the high correlated flux level around zero OPD, EWS works best if the OPD scan of the interferometric measurements is offset by about $5$\,$\lambda$ from zero OPD for the low spectral resolution of the prism. Although we used both methods independently to confirm the visibility results, we prefer the coherent method that is very sensitive in detecting weak fringe signals for our case of a faint source. After detailed tests for our case (faint source observed under good weather conditions), we chose to increase the sensitivity further by smoothing in the frame domain with a Gaussian width of 15 (parameter {\it gsmooth}, set by default to 4) before determining the group delay. Tests with the temporal high-pass smoothing width (parameter {\it smooth}) introduced to suppress background fluctuations showed no improvements when using other values than the default 50.

The resulting calibrated visibilty, derived by averaging the ratios of the raw visibilities of TW~Hya and the corresponding instrumental visibilities determined by observing HD~95272, is shown in Fig.~\ref{fig1}. The errors are mainly contributed by the uncertainties in the interferometric measurements of TW~Hya and not by those of the transfer function. Also plotted in Fig.~\ref{fig1} are the instrumental visibilities determined by the calibrator measurements obtained on the same baseline in that night. The rms scatter of these calibration measurements ranges from $\sim 0.02$ between 10\,$\mu$m and 11\,$\mu$m to $\sim 0.05$ at 8\,$\mu$m and 13\,$\mu$m. The feature in the wavelength range between 9.3\,$\mu$m and 9.9\,$\mu$m results from the ozone band of the earth's atmosphere that leads to a reduced count rate at these wavelengths. It is also clearly visbile in the correlated flux (Fig.~\ref{fig3}). Unfortunately, the extinction is very sensitive to the airmass so that a proper normalisation of the visibility according to Eq.~(\ref{eq1}) is very problematic in this wavelength range.

\subsection{Photometry\label{drphot}}

The photometric fluxes $F^{\rm total}\left(\lambda\right)$ for the visibility calculations are derived by subtracting the background by means of chopping techniques. However, in particular for faint sources, a second step of background subtraction is required for obtaining a reliable spectrophotometry. For this we performed a linear fit of the off-source background residuals for each wavelength and then subtracted this component from the chopping results.

The absolute spectra $F^{\rm abs}\left(\lambda\right)$ can now be derived from $F^{\rm total}\left(\lambda\right)$ by following the standard procedures applied when reducing mid-infrared spectra. The different airmasses of the scientific target and the spectrophotometric calibrator HD~95272 and the temporal variations in the seeing conditions have to be taken carefully into account. For the flux calibration we utilised the absolute fluxes for HD~95272 as published by \cite{cohen99}. 

The resulting spectrum is plotted in Fig.~\ref{fig2} and the Keck data from \cite{weinberger02} are presented for comparison. Both spectra agree very well both in shape and in the absolute fluxes. Also the Spitzer spectrum of TW~Hya (Sect.~\ref{res}) is in very good agreement. By either increasing the Spitzer spectrum or reducing the MIDI spectrum by 10\%, the two spectra fall on top of each other. Interestingly, this relative error is what we estimate for the absolute calibration error of the Spitzer spectrum.

\subsection{Correlated flux\label{drcorr}}

The wavelength-dependent correlated flux $F^{\rm corr}\left(\lambda\right)$ can now be derived by a simple multiplication of the quantities determined above:
\begin{equation}
F^{\rm corr}\left(\lambda\right) = V\left(\lambda\right)\cdot F^{\rm abs}\left(\lambda\right) 
\mathrm{\, .}\label{eq2}
\end{equation}
On the other hand, one may think of calibrating $F^{\rm corr}\left(\lambda\right)$ directly from the interferometric measurements instead. When comparing the two methods, we found in the case of TW~Hya that the resulting correlated spectra are very similar, but that the error bars are larger when deriving the correlated flux from the already calibrated quantities $V\left(\lambda\right)$ and $F^{\rm abs}\left(\lambda\right)$. These increased errors simply reflect that the determination of the correlated flux by using Eq.~(\ref{eq2}) requires five independent quantities more than the direct method.

The directly calibrated correlated flux is plotted in Fig.~\ref{fig3}. The silicate feature looks flatter and broader than in the total flux, suggesting we see larger particles here and more crystalline dust. A detailed analyis is given in Sect.~\ref{composition}.

%

\section{Spectrophotometry with the Spitzer satellite\label{res}}

The presented Spitzer spectra are based on the intermediate data products (low-resolution: {\it droopres}, high-resolution: {\it rsc}) processed through the Spitzer data pipeline (version S13.2.0). Partially based on the {\it SMART} software package \citep{higdon04}, these intermediate data products were further processed using spectral extraction tools developed for the `Spitzer Legacy Science Programs' `Formation and Evolution of Planetary Systems' (FEPS) \citep{meyer06} and `from Cores to Disks' (C2D) \citep{evans03}. For the low-resolution observations with the short wavelength module ($5-14$\,$\mu$m), the spectra were extracted using a 6.0 pixel fixed-width aperture in the spatial dimension. The background was subtracted using associated pairs of imaged spectra from the two nodded positions along the slit. This also subtracts stray light contamination from the peak-up apertures and adjusts pixels with anomalous dark current relative to the reference dark frames. Pixels flagged by the `Spitzer Science Center' (SSC) pipeline as `bad' were replaced with a value interpolated from an 8-pixel perimeter surrounding the bad pixel.

For the high-resolution spectra, all `Basic Calibrated Data' (BCDs) from both dither positions were combined within the extraction algorithm. The extraction was performed by integrating over a source profile fit in the cross-dispersion direction along the slit. The source profile is a template created from standard-star observations (including sky measurements). The width and centre of this template were adjusted for TW~Hya to encompass 95\% of the observed flux. This fitting process was performed by using the highest quality data. Once the source profile was fitted, a (uniform) local-sky background level, which is wavelength-dependent, was estimated. This method also reduces the effects of unidentified bad pixels. 

Both the low- and high-resolution spectra were calibrated using a spectral response function derived from IRS (`InfraRed Spectrograph') spectra and Cohen or MARCS stellar models for a suite of calibrators provided by the SSC. To remove any effect of pointing offsets, we matched orders based on the point spread function of the IRS instrument, correcting for possible flux losses. Offsets can also result in low-level fringing at all wavelengths in the high-resolution spectra. We removed these fringes using the {\it irsfringe} package developed by \cite{lahuis03}. We estimate that the relative flux calibration error across orders is 1\% and the absolute calibration error 10\%.

\begin{table*}
\caption{Overview of dust species used. For each component its lattice structure (A: amorphous, C: crystalline), chemical composition, and shape as well as a reference to the laboratory measurements of the optical constants are specified.}
\label{table5}
\centering                        
\begin{tabular}{lllcllc}
\hline\hline
\noalign{\smallskip}
\# &  & Species 				    & Structure & Chemical Formula\hspace{0.7cm} & Shape\hspace{0.7cm} & Reference\\
\noalign{\smallskip}
\hline
\noalign{\smallskip}
1  &  & Amorphous silicate (Olivine stoichiometry)  & A     & MgFeSiO$_{4}$	   & Homogeneous   & (1)\\
2  &  & Amorphous silicate (Pyroxene stoichiometry) & A     & MgFeSi$_{2}$O$_{6}$ & Homogeneous   & (1)\\
3  &  & Carbons 				    & A     & C		   & Homogeneous   & (2)\\
4  &  & Silica  				    & A     & SiO$_{2}$	   & Inhomogeneous & (3)\\
5  &  & Forsterite				    & C     & Mg$_{2}$SiO$_{4}$   & Inhomogeneous & (4)\\
6  &  & Clino Enstatite 			    & C     & MgSiO$_{3}$	   & Inhomogeneous & (5)\\
\noalign{\smallskip}
\hline
\noalign{\smallskip}
\multicolumn{7}{l}{\parbox{17.6cm}{References: (1) \cite{dorschner95}; (2) \cite{preibisch93}; 
(3) \cite{henning97}; (4) \cite{servoin73}; (5) \cite{jaeger98}}}\\
\noalign{\smallskip}
\hline
\end{tabular}
\end{table*}

%

\section{The composition of the circumstellar material\label{composition}}

\subsection{The dust composition\label{dc}}

\begin{figure}
\includegraphics[height=8.5cm,angle=90]{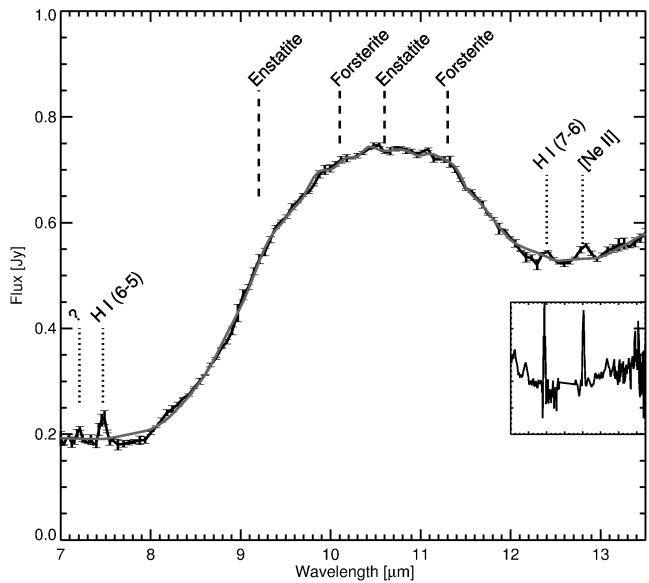}
\caption{Spitzer IRS observations of TW Hya in the 10\,$\mu$m wavelength region. Shown is the low-resolution IRS spectrum (black) between 7 and 13.5\,$\mu$m with the best-fit dust model described in Sect.~\ref{dc} overplotted (grey). As an inset the identically scaled high-resolution spectrum is shown between 12 and 13.5\,$\mu$m. The flux-level of both spectra is the same, but the inset has been shifted down for clarity. Also indicated in the figure are the positions of the main crystalline silicate bands and gas lines.}
\label{fig4}
\end{figure}

\begin{table}
\caption{Results of the spectral decomposition of the low-resolution Spitzer spectrum with a continuum temperature T$_c$=1500K and a dust temperature T$_d$=235K.}             
\label{table2}      
\centering                        
\begin{tabular}{lccccc}        
\hline\hline                 
\noalign{\smallskip}
&&&\multicolumn{3}{c}{Mass fraction [\%]}\\
Component &  &  & \ \ 0.1~$\mu$m\ \ & \ \ 1.5~$\mu$m\ \ & \ \ 6.0~$\mu$m\ \ \\    
\noalign{\smallskip}
\hline
\noalign{\smallskip}
Amorphous Olivine  &  &  & 66 & -- &  2 \\
Amorphous Pyroxene &  &  & 18 &  3 & -- \\
Amorphous Carbon   &  &  & -- & -- &  4 \\
Silica             &  &  & -- & -- & -- \\
Forsterite         &  &  &  3 & -- & -- \\
Enstatite          &  &  &  4 &  1 & -- \\
\noalign{\smallskip}
\hline                                  
\end{tabular}
\end{table}

To characterise the composition of the circumstellar dust we analysed the Spitzer low-resolution spectra in the 10\,$\mu$m wavelength region; see Fig.~\ref{fig4} and \cite{uchida04}. For this we applied a dust model and analysis method for determining the physical properties of the dust grains contributing to the observed emission similar to those successfully applied in previous studies of the 10\,$\mu$m region \citep{vanboekel05}. Though dust grains in protoplanetary disks are most likely highly irregular aggregates containing many different dust constituents, we assume that the aggregates are extremely porous and that the individual constituents making up the aggregate may interact with the radiation field as separate entities, as in the case of interplanetary dust particles \citep{henning96}. Therefore, we assume that the observed emission can be represented by the sum of the emission of individual dust species. Table~\ref{table5} summarises the dust species used in our analysis, including all dust species commonly identified in protoplanetary disks, e.g.~\cite{sargent06}, \cite{vanboekel05}, \cite{bouwman01} or \cite{bouwman07}. 

Crystalline silicates such as forsterite (Mg$_2$SiO$_4$) and enstatite (MgSiO$_3$) have many strong and narrow resonances in the wavelength range covered by the IRS spectrograph. These rotational/vibrational bands allow for an accurate determination of the chemical composition, grain size, and structure of the crystalline dust species. We find, like in the above-mentioned studies, that the pure magnesium end-members of the olivine and pyroxene families, together with the assumption of an inhomogeneous grain structure and non-spherical grain shape, give the best match to the observed spectral features. Here we use a distribution of hollow spheres \citep{min05}, which gives an excellent match to the observed band positions and shapes.

For the amorphous iron/magnesium silicates the exact composition and grain shape cannot be constrained as well as for the crystalline silicates. Models of the broad amorphous silicate resonance at $\sim 10$\,$\mu$m  are to some extent degenerate. Silicate models with varying magnesium-to-iron ratios and different grain shapes/homogeneity can reproduce the observed bands, e.g.~\cite{min07}. To allow for a direct comparison with previous studies, we assume here that the amorphous silicates have an equal magnesium-to-iron atomic ratio and stoichiometries consistent with olivine and pyroxene. Further, for the amorphous iron-magnesium silicates we assume homogeneous, spherical grains for which we calculate the absorption coefficients using Mie theory, e.g.~\cite{bohren83}, and the measured optical constants of these materials (see Table~\ref{table5} for references). 

To take the effect of grain size into account, we calculated for each of the different grain species the opacities for three different grain sizes of 0.1\,$\mu$m, 1.5\,$\mu$m, and 6.0\,$\mu$m, respectively. These sizes sample the range of spectroscopically identifiable grain sizes at infrared wavelengths at the SNR of our data. 

For the analysis of the dust composition we fitted the following emission model to the Spitzer low-resolution spectra by using a linear least-square minimisation:
\begin{equation}
\label{eq3}
F_\nu = B_\nu(T_\mathrm{cont})\, C_0 + B_\nu(T_\mathrm{dust}) \left(\sum_{i=1}^3\sum_{j=1}^6 C_{i,j}\,
\kappa_\nu^{i,j} \right),
\end{equation}
where $B_\nu(T_\mathrm{cont})$ denotes the Planck function at the characteristic continuum temperature $T_\mathrm{cont}$, $B_\nu(T_\mathrm{dust})$ the Planck function at the characteristic grain temperature $T_\mathrm{dust}$, $\kappa_\nu^{i,j}$ is the mass absorption coefficient for species $j$ (six in total) and grain size $i$ (three in total), and $C_0$ and  $C_{i,j}$ are the weighting factors of the continuum and the dust components, respectively.

Figure~\ref{fig4} shows the resulting best-fit model to the Spitzer spectrum. As one can see, an excellent agreement between model and observations can be obtained using the dust species listed in Table~\ref{table5}, without the need for additional components. However, the degeneracy between the contribution of the carbons and that of the continuum cannot finally be solved. The resulting model parameters are listed in Table~\ref{table2}. The mid-infrared emission is dominated by small amorphous silicate grains. The mass fraction of crystalline grains is about 8\%. In the modelling of the disk we adopt the dust composition as listed in Table~\ref{table2}.

\subsection{Gas composition}

Apart from the emission features from dust grains, we can also observe gas emission lines in the Spitzer spectra (Fig.~\ref{fig4}). In the low-resolution spectra emission lines at 7.47, 12.38 and 12.81\,$\mu$m can be detected. We identify these lines with emission from H\,I (6-5), H\,I (7-6), and [Ne\,II], respectively. The last two lines can also be observed in the high-resolution spectrum (see inset Fig.~\ref{fig4}). 


\cite{pascucci07} observed a number of optically thick dust disks in systems having similar accretion rates and X-ray luminosities to TW~Hya. They detected [Ne\,II] and HI\,(7-6) toward one transition disk and [Ne\,II] alone from three other disks that have reduced mid-infrared emission in comparison to classical T~Tauri star disks. The total line luminosity log L$_{\rm{[NeII]}}$ = 28.25\,erg/s measured with Spitzer for TW~Hya is similar to the values observed in the sources studied by \cite{pascucci07}. They argue that the [Ne\,II] emission most likely originates from neon atoms at the disk surface layer ionised by stellar X-ray or extreme UV photons. The [Ne\,II] line traces a gas mass of about $10^{-9}\,\mathrm{M}_{\odot}$, which could either be in the disk surface layer or the optically thin, depleted inner region (see Sect.~\ref{model}) of the disk. Unfortunately, the resolution of MIDI is too low to detect the line clearly in the correlated spectrum and thus to compare its luminosity in the innermost parts of the disk with the value derived from the total Spitzer spectrum. However, recent measurements obtained with `Michelle' at Gemini spectrally resolved the [Ne\,II] line and finally revealed its origin (Pascucci, private communication). On the other hand, the observed line luminosities of log L$_{\rm{HI(6-5)}}$ = 28.87\,erg/s and log L$_{\rm{HI(7-6)}}$ =  28.24\,erg/s are too strong to be explained by an ionised disk surface or magnetospheric accretion flows. The H\,I lines are most likely originating from an accretion shock or the stellar corona. ROSAT and XMM-Newton observations \citep{stelzer04, robrade06}, which can be explained well by X-ray emission from a metal depleted accretion shock, seem to support this hypothesis.

\subsection{Location of the dust species}

As shown by \cite{vanboekel04}, the correlated and uncorrelated fluxes provided by MIDI can be used to constrain the radial distribution of the emitting silicate grains. When comparing the correlated spectrum (Fig.~\ref{fig3}) with the total spectrum (Fig.~\ref{fig2}), it becomes obvious that the overall amorphous silicate band is absent in the correlated spectrum. This indicates that the region populated by the small amorphous grains can be resolved by MIDI and thus is localised at distances larger than a few AU from the central star.

\begin{figure}
\includegraphics[height=8.5cm,angle=90]{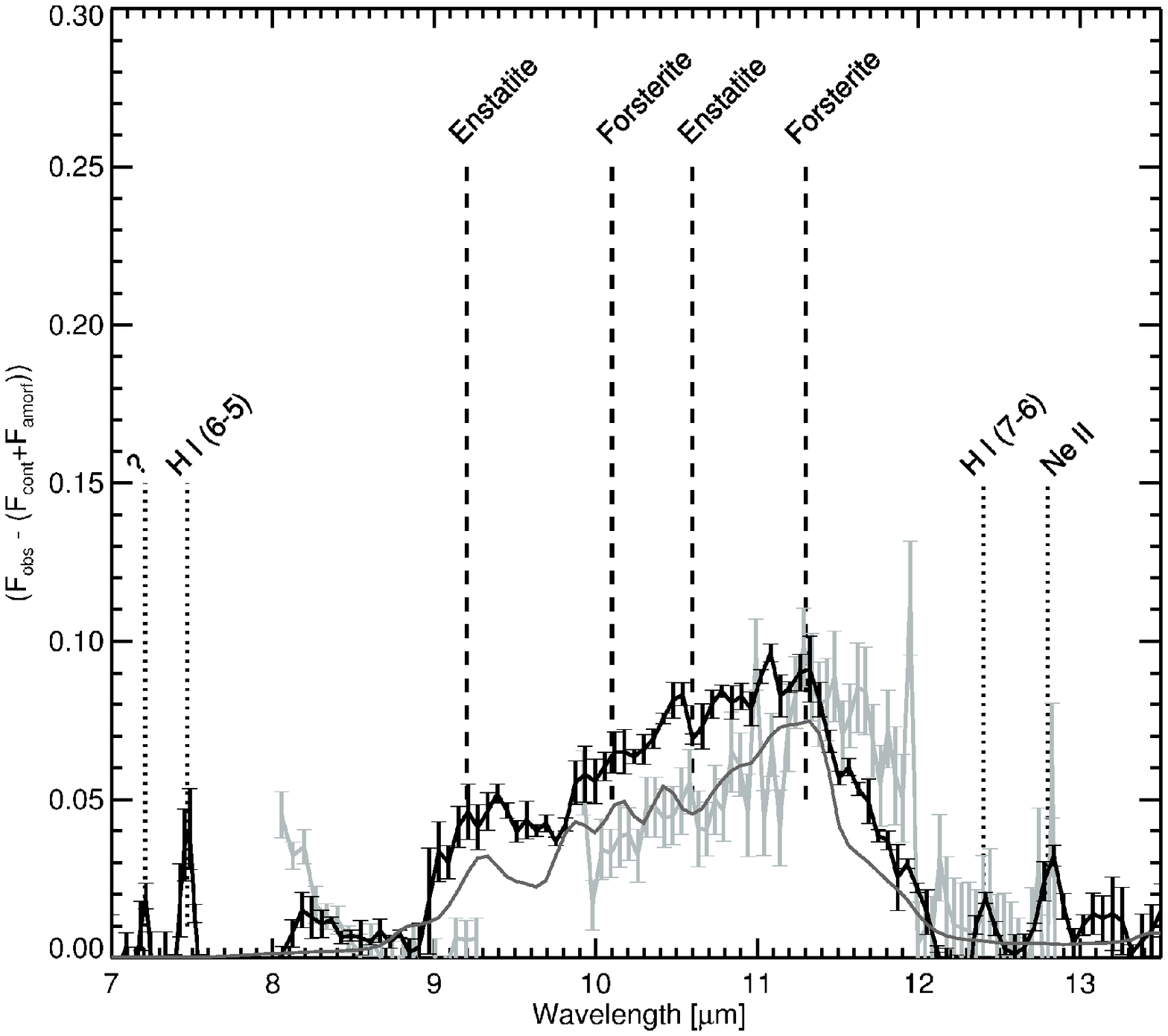}
\caption{\label{crystalline_emission}Comparison of the continuum subtracted correlated flux (light grey, wavelength range of the ozone band ignored) with the residual Spitzer spectrum after subtraction of the best-fit continuum and amorphous silicate model (black). Also plotted in this figure is a crystalline silicate model of the MIDI continuum-subtracted correlated spectrum (dark grey). Like in Fig.~\ref{fig4}, the positions of the main crystalline silicate bands and the observed gas lines have been indicated.}
\label{fig5}
\end{figure}

However, a weak remnant of the emission band can be seen in the correlated flux. By fitting a first-order polynomial between 8.5 and 12.5 to the correlated spectrum to estimate the continuum emission, one is left with the residual emission spectrum shown in Fig.~\ref{fig5}. When comparing this to the Spitzer spectrum after subtracting from the latter the fitted amorphous and continuum contribution, i.e.~taking only the infrared emission from the crystalline silicates into account, the match is already very good. This shows that the crystalline silicates are located much closer to the star than the amorphous species and that the emitting region can only be resolved marginally with our interferometric observations. The absence of crystalline emission signatures in the Spitzer spectra at wavelengths longwards of the N-band is another hint of this finding (see Sect.~\ref{lcg}). 

Unfortunately, the reduced spectral resolution and the worse signal-to-noise ratio of the MIDI spectrum with respect to the Spitzer result allow no meaningful spectral decomposition of the MIDI residual spectrum. However, a comparison shows that the correlated flux contains at least two thirds of the crystalline silicates we see in the Spitzer spectrum and an enstatite fraction that is lower by a factor of two (see Fig.~\ref{fig5}). This finding seems to indicate that the enstatite grains have a slightly larger spatial extension. But this could also be an artifact of the very simple continuum subtraction and the much larger uncertainties in the MIDI observations compared to the Spitzer spectrum.

%

\section{Model\label{model}}


The stellar parameters of TW~Hya are given in Table~\ref{table3}. For a more realistic modelling approach we used a Kurucz model \citep{kurucz92} for a star with 4000\,K, a surface gravity of $\log g [\mathrm{cm}\,\mathrm{s}^{-1}]=4.5$, and solar metallicity at the given distance instead of a mere blackbody. Although this template does not exactly fit the properties of TW~Hya, it is a reasonable approach that has also been used by \cite{eisner06}.

\begin{table}[t!]
\caption{Stellar parameters for TW~Hya}
\label{table3}
\centering                        
\begin{tabular}{lcl}
\hline\hline
\noalign{\smallskip}
                  &                                    & Reference\\
\noalign{\smallskip}
\hline
\noalign{\smallskip}
spectral type     & \hspace{1.7cm}K7V\hspace{1.7cm}     & (1)\\
T$_{\rm eff}$     & 4000\,K                             & (2)\\
mass              & 0.6\,M$_{\odot}$                    & (2)\\
luminosity        & 0.19\,L$_{\odot}$                   & (3) + parallax\\
radius            & 0.91\,R$_{\odot}$                   & calculated\\
age               & $5-15$\,Myr                         & (4)\\
accretion rate    & 4$\times$10$^{-10}$\,M$_{\odot}$/yr & (5)\\
extended flux     & 7\% at 2\,$\mu$m                    & (6)\\
distance          & $51\pm4$\,pc                        & (7)\\
A$_\mathrm{V}$    & 0.0\,mag                            & (1)\\
\noalign{\smallskip}
\hline                 
\noalign{\smallskip}
\multicolumn{3}{c}{\parbox{8.4cm}{References: (1) \cite{rucinski83}; (2) \cite{lb82}; (3) \cite{calvet02}; 
(4) \cite{uchida04}; (5) \cite{muzerolle00}; (6) \cite{johns01}; (7) \cite{mamajek05}}}\\
\noalign{\smallskip}
\hline                 
\end{tabular}
\end{table}

\subsection{The circumstellar disk}

We describe the disk using a model for passive irradiated flared disks based on the ideas of \cite{chiang97}. In this picture the dusty part of the disk can be described with two layers: a disk interior layer and a disk surface layer. The flared shape of the disk ensures that the stellar light can irradiate and heat the surface layer to the typical temperature of optically thin dust. The absorbed energy is then radiated away by the layer: half of this energy is radiated away from the disk and is then observable as optically thin warm dust radiation with solid state features in emission. The other half is emitted down into the disk interior, where it is subsequently re-emitted as optically thick featureless emission at wavelengths longer than the emission from the surface layer. 

\begin{figure}[h!]
\begin{center}
\includegraphics[width=8cm]{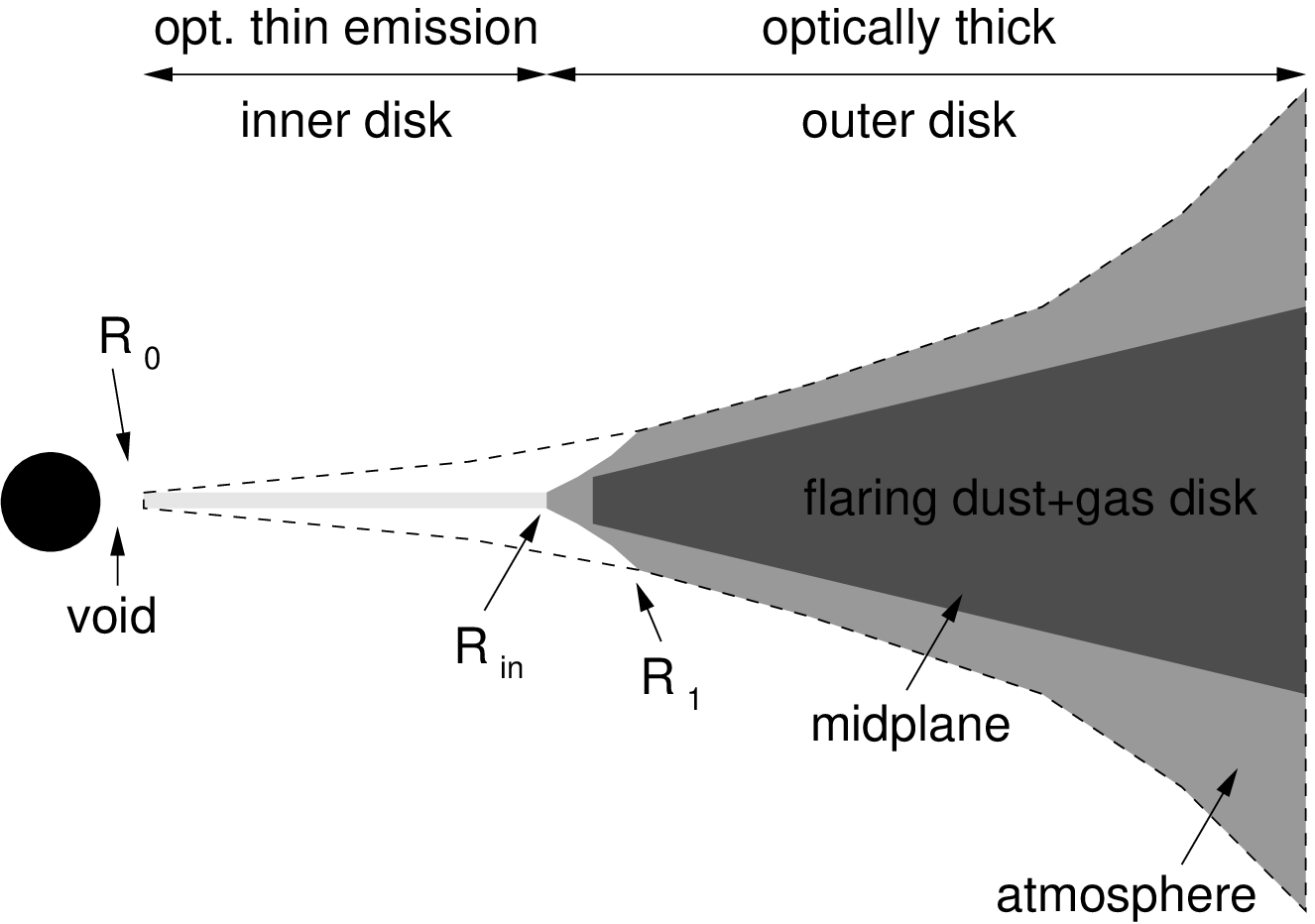}
\end{center}
\caption{Sketch of the geometry of the disk and how the inner dust rim is modelled. A vertical cross-cut of the disk is shown.}
\label{fig6}
\end{figure}

\begin{figure*}
\includegraphics[height=18cm,angle=90]{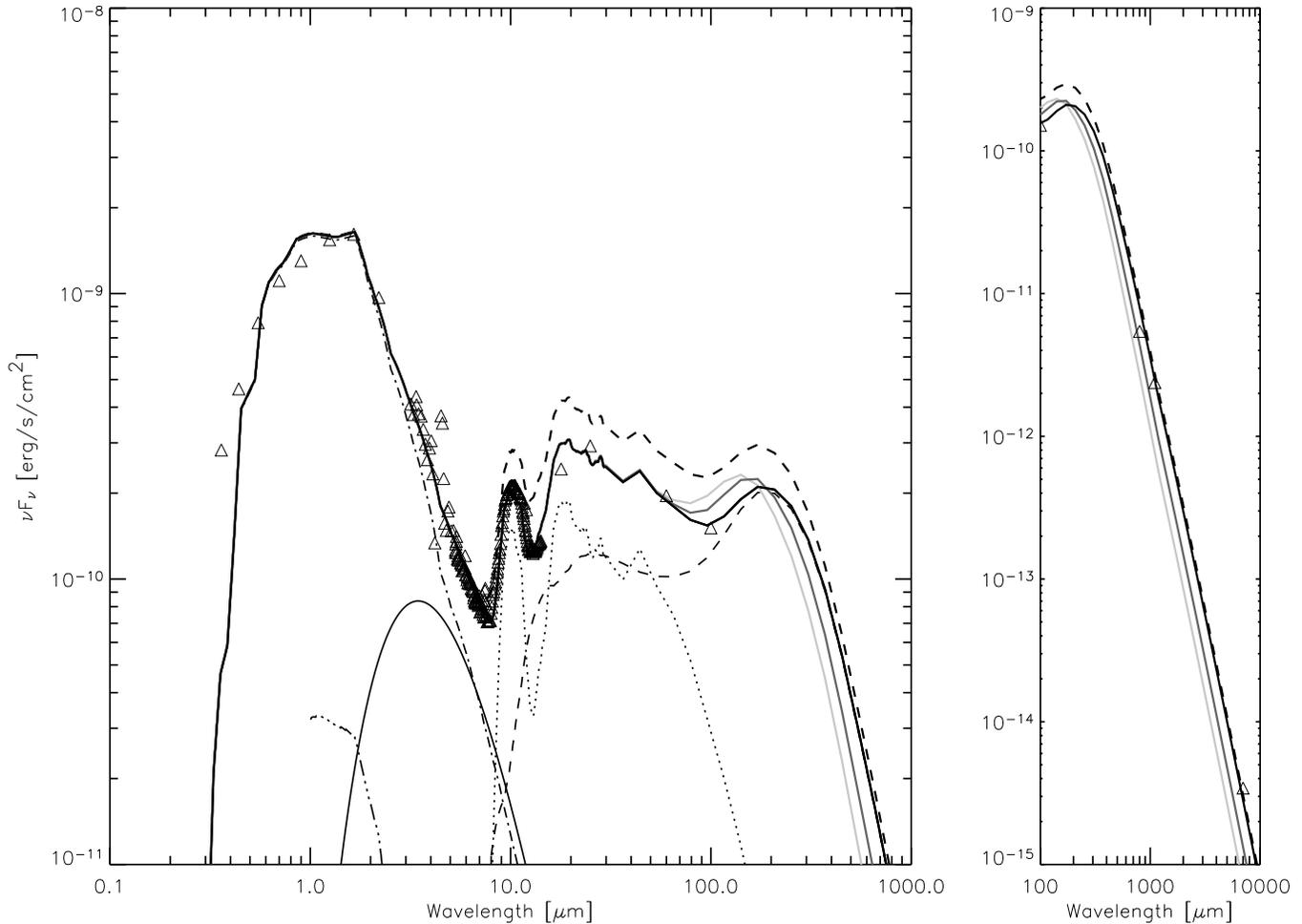}
\caption{{\it Left Panel:} Comparison of the observed SED (triangles) and the model fit (black) described in the text. The total SED (solid thick) consists of the contribution from the star (dashed-dotted), the inner optically thin disk (solid thin), the midplane of the optically thick outer disk (dashed), and its atmosphere (dotted). The additional component in the near-infrared is the contribution due to scattering. To demonstrate the influence of the disk mass, the SEDs of disks with a mass of 0.16\,$M_{\odot}$ (black), 0.06\,$M_{\odot}$ (dark grey), and 0.03\,$M_{\odot}$ (light grey) are given after applying a sedimentation parameter of 0.70 (black), 0.73 (dark grey), and 0.75 (light grey) to scale the mid-infrared regime. When in the model fit (thick solid) the sedimentation is no longer taken into account, the SED is overestimated in the mid-infrared by a factor of $\sim 1.5$ (thick dashed), but the visibility reproduces the short wavelengths very well (see Fig.~\ref{fig9}). {\it Right Panel:} SED longward of 100\,$\mu$m for the four scenarios described above. {\it Measurements:} The observational data are taken from \cite{mekkaden98} for the visual, \cite{rucinski83} and \cite{sitko00} for the near-infrared, this paper for the mid-infrared, \cite{weinberger02}, the IRAS data archive, and \cite{weintraub89} for the sub-millimetre, and \cite{wilner00} for the millimetre wavelength-regime$^5$.}
\label{fig7}
\end{figure*}

Most of the details of the model used here are described in the paper by \cite{dullemond01}. However, we treat the inner rim differently. In the \cite{dullemond01} model the rim was included as a vertical wall. The work of \cite{isella05} has shown that these rims are not vertical, but strongly rounded-off, and recently interferometric results could be reproduced successfully with this model \citep{isella06}. This finding has strong consequences for the inclination dependence of the infrared flux and even stronger consequences for the infrared interferometric measurements. We therefore include a sort of `rounded-off' rim in our model in a similar way to what is described by \cite{calvet02}: normally the pressure scale height of the disk $H_p$ is calculated self-consistently using the \cite{chiang97} approach. Close to the inner rim $R_{\rm in}$, however, we artificially reduce $H_p$ smoothly, such that it reaches $0$ at $R_{\rm in}$. The width $\xi$ in which this reduction is done -- from no reduction at $R_1\equiv (1+\xi)\,R_\mathrm{in}$ to full reduction at $R_{\mathrm{in}}$ -- sets the sharpness of the rim. If $\xi=0.1$, the rim is rather sharp, while if $\xi=1$ the transition from no dust to normal flaring is very gradual. Rather than attempting to calculate $\xi$ self-consistently from the physics of possible mechanisms that might smooth-out the rim, we simply set $\xi$ and consider it a free parameter. The precise way in which we reduce $H_p$ between $R_{\mathrm{in}}$ and $R_1$ is also ad-hoc: if we call $\psi$ the reduction factor, i.e.~$\hat{H}_p(R) =\psi(R)\,H_p(R)$, it is defined as
\begin{equation}
\psi(R)\equiv\left({R-R_{\mathrm{in}}}\over{R_1-R_{\mathrm{in}}}\right)^q\mathrm{,}\label{eq4}
\end{equation}
\footnotetext[5]{With the exception of the 18\,$\mu$m measurement, these data are identical to those used by \cite{calvet02} and \cite{eisner06} for their SED. In the Q-band we substitute the result of \cite{jayawardhana99} by the more recent measurement published in \cite{weinberger02}.} 
where $q$ is again a parameter we set freely. Therefore, this approach enables us to investigate a variety of modifications of the geometry of the rim by just setting the two independent parameters $\xi$ and $q$. Since we are mainly interested in this geometry and a direct comparison with the model of \cite{calvet02}, we are confident that this procedure is reasonable, especially when taking into account that we have only one independently measured visibiliy curve. A pictographic representation of the model is given in Fig.~\ref{fig6}. 

After predicting the pressure scale height of the disk, the surface height of the disk can then be computed from the pressure scale height when the surface density and dust opacity is known and if the assumption is made that the dust and the gas are well mixed. As noted by \cite{chiang01} dust sedimentation can lower the surface height for a given pressure scale height. They modelled this by introducing a reduction parameter that reduces the surface height somewhat from the computed value. In our model we also allow a (global) parameter that almost directly scales the flux from the disk and thus the SED (Fig.~\ref{fig7}), but has only very little influence on the brightness distribution and the visibility (Fig.~\ref{fig9}), respectively. 

For the outer dusty part of the disk, we use the mixture of dust opacities given in Table~\ref{table2} derived by modelling the Spitzer data (Sect.~\ref{composition}). Intrinsically, there is no treatment for scattering in the model: only absorption and re-emission is included. The total abundance of dust is taken to be 0.01 by mass with respect to the gas.

\subsection{Scattering}

Despite the fact that the Chiang-Goldreich model described above ignores scattering, this effect is, according to Mie theory, proportional to $\lambda^{-4}$ and hence may play an important role at visual and near-infrared wavelengths. For the discussion of extended emission in the K-band, we thus include a simple model of the additional contribution of scattered light to the near-infrared emission (Appendix~\ref{sctt}). 

We find that this contribution is only a small effect (see Fig.~\ref{fig7}) and reduces the visibility prediction for TW~Hya in the K-band by about 0.01. However, for a fully consistent model, the heating losses caused by the scattering would have to be taken into account by the Chiang-Goldreich model.



\section{Fitting SED and Visibility\label{sed_visibility}}

\subsection{Previous models}

In the mid-infrared region, where the interferometric measurements on the VLTI contribute directly, the SED of TW~Hya (Fig.~\ref{fig7}) is characterised by a strong silicate emission, by quite notable excess emission in the range of $3-8$\,$\mu$m, and by a strong increase in F$_{\nu}$ towards the $20-25$\,$\mu$m range. \cite{calvet02} argue that they cannot obtain the strong silicate emission from a disk atmosphere, if they do not neglect scattering. They put the edge of the optically thick disk at $3-4$\,AU, where it does not contribute to the 10\,$\mu$m emission. The silicate emission in their model is due to an optically thin distribution of micron-sized particles, populating a geometrically rather thin volume in the inner disk with constant surface density, which adds sufficiently to the emission longward of the silicate band. The 2\,$\mu$m interferometry of TW~Hya by \cite{eisner06} added the constraint that much of the $7\pm4$\% of extended emission observed at that wavelength \citep{johns01} must not come from close to the evaporation limit ($R_0\sim 0.025$\,AU), since otherwise the visibility could not be as low as their observed value of 0.88 $\pm$ 0.05. They modified the optically thin part of the \cite{calvet02} model, so that the surface density reflects a radial distribution according to a power law of 1.5 and the particles have an opacity proportional to $\lambda^{-1}$ to create both sufficient 2\,$\mu$m radiation and the $3-8$\,$\mu$m near-infrared excess. They added a dust-free hole inside 0.06\,AU, which is required to reproduce the comparatively low K-band visibility. Since they did not try to reproduce the SED beyond 8\,$\mu$m, the 10\,$\mu$m range, where we measured visibilities, is not covered by their model.

Therefore, we only have to check the model of \cite{calvet02} for compatibility with the new N-band interferometric data. Lacking their original model intensity data, we simulated them with a Chiang-Goldreich model, keeping position and shape for the inner edge of the optically thick disk ($3-4$\,AU), constant optical thickness, temperature law, and particle size range ($0.9-2.0$\,$\mu$m) for the optically thin inner disk as given in their paper. We assumed the dust composition to be `astronomical silicate' \citep{laor93} with a 7\% addition of carbon for both parts of the disk. The SED predicted from this coarse remake looked to be reasonably reproduced and, after adjustment of the free parameters of our model (see Table~\ref{table4}), fitted the observations well, except for wavelengths longer than 100\,$\mu$m. We therefore consider this simulation as a valid representation of the model of \cite{calvet02} for the mid-infrared range around 10\,$\mu$m. The visibilities corresponding to this model are compared to the observations in Fig.~\ref{fig9}. Clearly, the predicted values are far too high and show the wrong trend with wavelength. To a similar degree the model overpredicts the visibility observed at 2.14 $\mu$m by \cite{eisner06}. These results originate almost exclusively from the physically simple, easy-to-model, inner optically thin part of the model of \cite{calvet02}, which should be well represented by our simulation. In spite of its strength in modelling the outer disk, the model will apparently need substantial modification to accomodate the new interferometric data. Or the other way around, the visibility will help to decide whether a model that resembles the SED as well as the model introduced by \cite{calvet02} is more reliable.

\subsection{New mid-infrared constraints\label{mircon}}

The new constraints given by our 10\,$\mu$m interferometry of TW~Hya will now be used to derive a model of the circumstellar disk geometry for TW~Hya and, in this light, to consider the uniqueness of the two earlier conclusions about the disk of TW~Hya -- a dust-free zone within 0.06 AU and the limitation of the optically thick disk to radii beyond $3$ to $4$\,AU. We will concentrate on the inner 10\,AU or so of the circumstellar disk, since this is the region to which the data from MIDI refer. We do so by modelling the SED and the visibility with the help of the model introduced in the previous section. The dust is described by the composition derived with an isothermal fit to the silicate feature in Sect.~\ref{composition}. Realistic or not, even in our Chiang-Goldreich type model with a certain temperature distribution, this composition can well reproduce the shape of the silicate emission that is so central in our observed wavelength range (see Fig.~\ref{fig10}). However, we are aware of the approximative character of this model. 

\begin{figure}
\includegraphics[height=9cm,angle=90]{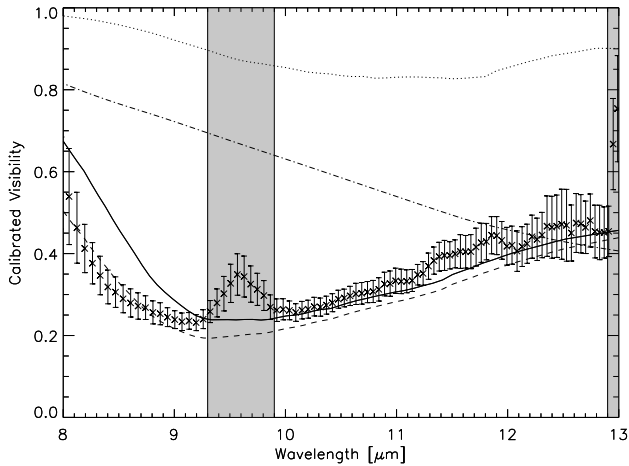}
\caption{Comparison of observed visibility and the best simultaneous fit (solid) to both SED and visibility. Without sedimentation the model overestimates the SED (see Fig.~\ref{fig7}), but especially at short wavelengths the visibility is very well reproduced (dashed). Likewise, the visibiliy found by reproducing the model described by \cite{calvet02} is shown (dashed-dotted), as well as that originating from a pure Chiang-Goldreich model fitting the SED too (dotted).}
\label{fig9}
\end{figure}

Also, we do not want to compete with the detailed argumentation in \cite{calvet02}, which has led to valuable physical insights into the outer disk regions. Especially, the explicit discussion therein about the disk mass cannot be taken up again, because in our dust composition mm- and cm-sized dust grains are not considered. These large particles contribute mainly to the flux in the sub-millimetre and millimetre wavelength regime and thus are not relevant to our mid-infrared analysis. However, when including the mm- and cm-sized dust grains, the high disk mass of 0.16\,$M_{\odot}$, which is now required to fit the fluxes longward of 100\,$\mu$m, would be reduced significantly. Figure~\ref{fig7} demonstrates the ambiguous nature of the mass as a fit parameter for the mid-infrared when adjusting the sedimentation. The differences between the various mass-sedimentation scenarios only play a role logward of 100\,$\mu$m. Moreover, the resulting mid-infrared visibilities are indistinguishable from one another (Fig.~\ref{fig9}). The visibility is thus not able to solve this ambiguity. Conclusions based on visibility measurements are basically geometric in nature and therefore qualitatively valid and reliable as far as questions of disk geometry are concerned.

In this vein let us investigate the observed visibility curve of TW~Hya. The in general low values ($\sim 0.3$ from $9$\,$\mu$m to $11$\,$\mu$m) mean that the typical size of the emitting  region -- characterised, e.g., by the diameter of a structure -- is of the order of $\lambda/2B$ to $\lambda/B$, where $B$ is the projected baseline of the interferometric measurement, or $1-2$\,AU. The increase at short wavelengths is a natural consequence of the fact that an increasing fraction of the emission comes from unresolved regions, including the star. Most informative is the rise of visibility with wavelength beyond 11 $\mu$m. This is the signature of an emitting region whose size varies little with wavelength, such as could be produced by a `cliff-like' inner edge of the optically thick outer part of the circumstellar disk. Indeed, any disk geometry compatible with the visibility measurements seems to require the presence of a `wall' at this inner edge with a diameter of about 1.5\,AU and with reduced emission inside. A Chiang-Goldreich-type model of the circumstellar disk with an inner radius similar to the sublimation radius, for example, immediately approaches the observed SED, but completely fails to reproduce the low observed visibilities (see Fig.~\ref{fig9}). The model presented by \cite{calvet02} also fails to reproduce both the shape and the low values of the measured visibility, although it includes the required `cliff-like' rim at an even larger distance from the central star than in our model. The reason for this finding becomes apparent when analysing the contributions of the various disk components to the SED. In the model developed by \cite{calvet02} the `wall' is so far away from the central heating source that it cannot contribute significantly to the N-band flux at all. In this wavelength range the visibility traces only the inner disk.

Although the large fraction of stellar luminosity that is reradiated at longer wavelengths leaves no doubt that the disk extending outside this circular wall is optically thick, the properties of the region inside this optically thick disk are not well-constrained. Nevertheless, if this inner region is responsible for the observed extended 2\,$\mu$m emission \citep{eisner06}, it cannot be empty. But what we assume there to produce the desired extended emission is largely arbitrary, as it has been in earlier models. The presence of accretion, though at a low level \citep{muzerolle00}, was used to justify the presence of optically thin material in this region showing either a) none (little) or b) pronounced silicate emission in the mid-infrared. 

The best combined fit to SED and visibility and our preferred solution is a model with an inner disk similar to that described in \cite{eisner06} and a transition region between the inner and the outer disk that is much closer to the star when compared with the results of \cite{calvet02}. The parameters of this model are given in Tables~\ref{table3} and \ref{table4}. It is compared to the observations in Figs.~\ref{fig7} and \ref{fig9}.

\begin{table}[t!]
\caption{Parameters for the circumstellar disk of TW~Hya}
\label{table4}
\centering                        
\begin{tabular}{lcl}
\hline\hline
\multicolumn{3}{c}{optically thick outer disk}\\
\hline
\noalign{\smallskip}
inclination             & \hspace{0.7cm}0$^{\circ}$ ($<$4$^{\circ}$)\hspace{0.7cm} & (1)\\
outer diameter          & 140\,AU		       & (2)\\
surface density         & 1$\times$10$^{4}$\,g/cm$^2$  & at 1 AU, fit parameter\\ 
power law               & $r^{-1.5}$  		       & (3)\\
sedimentation           & 0.7			       & fit parameter (see text)\\
mass                    & 0.16\,M$_{\odot}$	       & calculated (see text)\\
opacity                 & see text		       & this paper\\
\noalign{\smallskip}\noalign{\smallskip}
\hline\hline
\multicolumn{3}{c}{geometry of transition region, see Equation~(\ref{eq3})}\\
\hline
\noalign{\smallskip}
$R_{\rm in}$            & 0.5\,AU		       & fit parameter\\
$R_1$                   & 0.8\,AU		       & fit parameter\\
$\xi$                   & 0.6			       & calculated\\
$q$                     & 2.0			       & fit parameter\\
\noalign{\smallskip}\noalign{\smallskip}
\hline\hline
\multicolumn{3}{c}{optically thin inner disk}\\
\hline
\noalign{\smallskip}
$R_0$                   & 0.06\,AU		       & fit parameter\\
$T (R_0)$               & 1050\,K		       & fit parameter\\
surface density         & 9$\times$10$^{-7}$\,g/cm$^2$ & at 1 AU, fit parameter\\ 
power law               & $r^{-1.5}$  		       & fit parameter\\
mass                    & 1.2$\times$10$^{21}$\,g      & calculated\\
$\tau_\bot (2\,\mu m)$  & 0.06  		       & at $R_0$, calculated\\
$\tau_\bot (10\,\mu m)$ & 0.013 		       & at $R_0$, calculated\\
opacity                 & 10$^3$\,cm$^2$/g             & at 2\,$\mu$m, (4)\\
power law               & $\lambda^{-1.0}$             & (4)\\
\noalign{\smallskip}\noalign{\smallskip}
\hline
\noalign{\smallskip}
\multicolumn{3}{c}{\parbox{8.1cm}{References: (1) \cite{weinberger02}; (2) \cite{krist00}; (3) \cite{dallesio98}; (4) \cite{eisner06}}}\\
\noalign{\smallskip}
\hline                 
\end{tabular}
\end{table}

%

\section{Discussion\label{discussion}}

\subsection{The inner disk\label{id}}

We have no good reasoning for the character of the inner disk. We only know that its $3-8$\,$\mu$m emission must be substantially reduced with respect to a standard Chiang-Goldreich flaring disk model. 

Between the two possibilities considered by us, we obtain the best representation of the data by assuming an {\it optically thin distribution of particles showing the opacity given in \cite{eisner06}}, i.e.~proportional to $\lambda^{-1}$. In analogy to solar system interplanetary dust dynamics, we estimate that the Poynting-Robertson lifetimes even for 1\,mm particles at the transition region would be only $10^6$\,yr \citep{wyatt50}. For this model we therefore have to assume that accretion continues through the inner edge of the outer disk and that (smaller) particles showing features in the mid-infrared are removed on a much shorter timescale by a combination of typical sources also acting in interplanetary space: radiation and stellar wind pressure, magnetic acceleration, collisions, gas drag, and the Poynting-Robertson effect. All of these are more important for small particles.

Our second scenario is an {\it optically thin distribution of silicate dust showing pronounced silicate emission}. This does not produce a very good fit to the SED and visibility data because it tends to overpredict the 10\,$\mu$m emission. Moreover, the timescale for the Poynting-Robertson drag is proportional to the radius of the particle, which only makes this scenario feasible if there is continuous replenishment of these grains on short timescales.


However, both scenarios would face the same problem of reproducing the low observed K-band visibilities if the inner disk continues inward to the dust sublimation limit at $\sim 0.025$\,AU ($\sim 1500$\,K), because a too large fraction of the 2\,$\mu$m radiation would then be unresolved by the 62\,m baseline, i.e.~$\mathrm{u}=49.6$\,m and $\mathrm{v}=36.7$\,m, of the Keck observations \citep{eisner06}. Therefore, one is led to postulate, like \cite{eisner06} did, a truncation of the inner disk at $R_0\sim 0.06$\,AU or $\sim  12$ stellar radii, probably caused by the action of the stellar magnetic field, but somewhat outside the corotation radius of 6.3\,R$_*$ \citep{johns01}. Knowledge of the inner boundary of the inner disk may profit in the same way from spectrally dispersed near-infrared interferometry as the knowledge of the inner edge of the outer disk benefits from the interferometric MIDI observations in the mid-infrared. This complementary observation could already be performed with the sensitivity provided by AMBER on the VLTI \citep{petrov03} or the successfully tested grism mode of the Keck Interferometer \citep{eisner07}.

Qualitatively we thus reproduce the model results of \cite{eisner06}. Quantitatively, we have to assume a particle density that is higher by a factor of $\sim$ 1.5 to have the same infrared excess. However, our prediction for the squared visibility at a wavelength of 2.14\,$\mu$m and for the projected baseline length reported in \cite{eisner06} is 0.92 and thus consistent with the measurements at the Keck interferometer.

The {\it amount of diffuse 2\,$\mu$m emission} of $r =0.07\pm0.04$ \citep{johns01} can be kept in both scenarios discussed above, but the first one -- comparatively large particles -- allows the best overall representation. We therefore propose to consider this model for the inner disk as a serious possibility that deserves further investigation. We refer to the work by \cite{tanaka05} and note that \cite{ciesla07} discussed a mechanism that can lead to strong particle growth in the inner part of a circumstellar disk.

\subsection{Transition region}

The visibility measurements at 10 $\mu$m put the edge of the outer disk at $0.5-0.8$\,AU. This is much less than the radius of $3-4$\,AU for the inner disk derived by \cite{calvet02}. However, it remains the question what interaction causes this abrupt change between the outer and inner disk. \cite{calvet02} present arguments that a massive planet might be the underlying reason. They also mention GG~Tau and DQ~Tau as examples of where a companion has cleared the inner region of a larger disk -- documented by millimetre- and near-infrared-images of the GG Tau system -- and where accretion from the outside still occurs -- at least in the case of DQ~Tau. Although \cite{weinberger02} searched for such a companion to TW~Hya and excluded any point source brighter than a planet of 10\,M$_{\rm J}$ for separations $>50$\,AU from TW~Hya, this result is not directly relevant for our model, where the orbit of a planet or companion would have to be at $\sim 0.25$\,AU. \cite{herbig88} noted that three radial velocity determinations showed a spread of $\pm11$ km/s, making it worthwile to check for a close companion. A companion of 0.1\,M$_{\odot}$ at $0.20-0.25$\,AU from TW~Hya would have an orbital velocity of $43-48$\,km/s leading to a radial velocity of the primary of $0.5-0.6$\,km/s when assuming an orbital inclination of $4^{\circ}$ (see Table~\ref{table4}). This is much less than the scatter in these early observations. We are not aware of radial velocity monitoring of TW~Hya, although such a search could be fruitful. The low near-infrared brightness of a 0.1\,M$_{\odot}$ companion will make it difficult, but not impossible, to detect it even by an interferometric instrument like AMBER on the VLTI \citep{petrov03}. However, we emphasise that locally enhanced particle growth or photoevaporation of the dust can also lead to an optically thin inner disk.

According to our analysis, the abrupt change from the outer to the inner disk, which was the main feature found by \cite{calvet02}, occurs approximately by a factor of six closer to the star than previously concluded. The 10\,$\mu$m visibilities put hard constraints on the location of the inner edge of the outer disk. It happens that this new, smaller distance essentially coincides with the region of CO emission at about 0.5\,AU. This region was identified by \cite{rettig04} on the basis of the observed rotational temperature of CO transitions. The edge of the outer disk, if located at the small radius derived by us, would naturally qualify as the origin of the CO emission. \cite{calvet02} proposed the large radius of the optically thin inner disk due to their conclusion that the inner disk and not the upper layers of the optically thick outer disk produce the strong silicate emission feature. However, this solution of \cite{calvet02} is not unique but just one of the possibilities for describing the SED {\it unconstrained} by mid-infrared visibilities and the mm-images. We showed in Sect.~\ref{mircon} that, fully consistent with the mid-infrared visibilities, the silicate emission can be produced in the atmosphere of the outer optically thick disk when taking moderate sedimentation into account.

\subsection{Disk evolution\label{de}}

With its estimated age of $\sim 10$\,Myr, TW~Hya is among the oldest classical T~Tauri stars. We might expect that its circumstellar disk shows signs of evolution, like continuing crystallisation, growth of particles, beginning dispersal of the disk, and onset of planet formation.

The evidence is mixed. Given the large size of the circumstellar disk visible in scattered light \citep{weinberger02} and the strongly disk-dominated emission for wavelengths above 10\,$\mu$m, which led \cite{calvet02} to a model with a disk mass of 0.06\,M$_{\odot}$, the disk around TW~Hya is far from dispersal, even after 10\,Myr. \cite{calvet02} provide interesting suggestions to explain this unusual survival time. On the other hand, the existence of mm- to cm-sized particles, inferred from the slow fall of mm-emission with wavelength \citep{weinberger02, wilner05}, shows strong particle growth into a size-regime where dust settling to the midplane of the disk should become effective.

If most of the 10 $\mu$m emission comes from the upper disk layers, as we have argued, a much more primordial size distribution is present there. The mid-infrared measurements in general would not be sensitive to mm-sized particles, but the 10\,$\mu$m silicate emission feature would become broader and flatter in the presence of micron-sized particles. This occurs in many T~Tauri stars \citep{przygodda03}, but not in TW~Hya. Instead the emission profile here requires a vastly dominating fraction of submicron-sized silicate particles. Since general particle growth by coagulation would have not only led to an increase in the fraction of large grains in the midplane, but also in the disk atmosphere, our analysis indicates processes, e.g.~sedimentation, that have removed these particles. 

A second mode of dust evolution is crystallisation. Again, very little crystalline material contributes to the observed silicate emission from TW~Hya. In the interferometrically observed flux, representing emission from the innermost region with a diameter of a few AU, and probably coming from the inner edge of the disk, the situation is different. This is not totally unexpected since a similar behaviour has been found in Herbig Ae/Be stars \citep{vanboekel04}, but it is not obvious that the same has to hold for the much less luminous T Tauri stars. 

From these observations, it appears that both radial mixing that would distribute crystals and vertical mixing that would homogenise size distributions are not important in the disk of TW~Hya. For a deeper discussion of transport mechanisms it will be helpful to have the corresponding data available for a number of T~Tauri stars.

\subsection{Location of the crystalline grains\label{lcg}}

\begin{figure}
\includegraphics[height=9cm,angle=90]{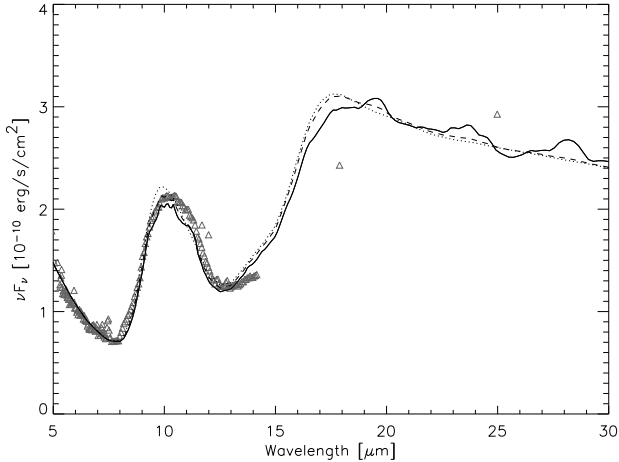}
\caption{Mid-infrared part of the SED for three different distributions of the crystalline grains: from the dust mixture found by fitting the Spitzer spectrum the crystalline grains have been removed nowhere (solid), outside the transition region (dashed), and in the entire disk (dotted). The underlying triangles represent the measured fluxes (for references see caption of Fig.~\ref{fig7}).}
\label{fig10}
\end{figure}

A direct indicator of an inhomogeneous distribution of the crystalline grains in the optically thick disk of TW~Hya are the silicate emission features at wavelengths longer than 20\,$\mu$m. These features are present in the model SED (Fig.~\ref{fig7}), but they are absent in the Spitzer spectrum. When removing from the entire disk the crystalline species, one finds the measured smooth SED at longer wavelengths, but the N-band that causes the dust composition of the model is no longer well represented (see Fig.~\ref{fig10}). Interestingly, when keeping the found dust mixture with its low percentage of crystalline grains only in the transition region, both the N-band emission and the featureless SED at longer wavelengths is reasonably reproduced.

Although there is evidence -- both from the correlated flux and the SED -- that the processed dust is close to the edge of the optically thick disk and may even be distributed throughout the optically thin inner disk, detailed investigations of the dust composition as a function of distance from the star are necessary for deeper insight into the location of the crystalline grains. We leave it to other studies to conclude on this.

\subsection{Millimetre maps}

The images obtained at 7\,mm by \cite{wilner00} are not discussed here. They are well reproduced by the outermost part of the disk modelled by \cite{calvet02}, which in principle could be joined to the modified inner parts of the disk, whose properties have been determined in this work with the help of mid-infrared measurements.

%

\section{Summary and conclusion\label{conclusion}}

To summarise and conclude:

\begin{itemize}
\item We were able to interferometrically measure the comparatively faint  T Tauri star TW~Hya. Similar studies will therefore be possible for a large number of low-mass young stars.
\item We confirm that the circumstellar disk of TW~Hya can be described best by a rather standard outer part with an abrupt transition to an inner part with reduced emission. However, based on 10\,$\mu$m visibility measurements, we put this transition region much closer to the star than it has been done in earlier models, namely between 0.5 and 0.8 AU. This transition region would qualify as the source of much of the CO emission observed by \cite{rettig04} to originate from a region around 0.5\,AU. 
\item The inner disk, as already discussed by \cite{eisner06}, does not appear to extend inwards fully to the dust sublimation and corotation limit. Spectrally dispersed, near-infrared interferometry may be needed to definitely clarify this point. 
\item Our results do not contribute anything new to the question of what causes the abrupt change in properties between the outer and the inner part of the disk. Possible answers are changing opacities induced by locally enhanced particle growth or a massive planet. However, in our proposed geometry, any potential companion or massive planet would have to revolve closer to the star and faster, and should thus be easier to detect by radial velocity monitoring.
\item Comparison of spectrophotometry from Spitzer with spectrally resolved correlated flux measurements from MIDI at the VLTI shows that most crystalline material appears to be concentrated within a few AU from the star. The disk of TW~Hya is not mixed well at the present epoch.
\item The parametrised models used in this work and in earlier papers are not completely satisfactory for application to an even modestly complex disk like the one of TW~Hya. A fully fledged radiative transfer/hydrodynamics code might give a physically more convincing representation of the now copiously available data relating to the geometry of this circumstellar disk.
\end{itemize}

%

\begin{acknowledgements}
This work is based in part on observations made with the Spitzer Space Telescope, which is operated by the Jet Propulsion Laboratory, California Institute of Technology under a contract with NASA. We thank the ESO staff at Paranal and Garching for their support during the preparation and execution of the observations. We thank A.~Schegerer for the initial modelling and the helpful discussions, and we are grateful to the anonymous referee for the very constructive comments. Th.~R.~cordially thanks `P{\"u}tz' for always accompanying him.
\end{acknowledgements}

\appendix

\section{Scattering\label{sctt}}

We calculate the contribution of light scattered at the optically thick part of the disk by
\begin{equation}
I_{scat}\left(\lambda\right) =F_*\left(\lambda\right)\tau\left(\lambda\right) A\left(\lambda\right) \Phi\left(\lambda,\vartheta\right)\mathrm{ ,}\label{eqa1}
\end{equation}
where  $F_*$ is the irradiation by the star at the location where the scattering occurs, $\tau\sim 1$ the optical thickness of the scattering surface, $A$ the albedo, and $\Phi$ the phase function taken from \cite{henyey41} as
\begin{equation}
\Phi\left(\lambda,\vartheta\right)={1-g\left(\lambda\right)^2\over 4^{\ }\pi} \left[1+g\left(\lambda\right)^2-2g\left(\lambda\right)\cos\vartheta\right]^{-3/2}\label{eqa2}
\end{equation}
with the average cosine of the angle of scattering $g$ and the scattering angle $\vartheta\sim 90^{\circ}$. The values for $A$ and $g$ are those for a mixture of uncoated submicron-sized silicate and graphite particles published by \cite{mathis83}. The foreground extinction has been assumed to be zero (see Table~\ref{table3}) at all wavelengths. This is sufficient for the near-infrared, but might be erroneous for shorter wavelengths. We further restricted the diameter of the emitting region to five times the Airy-disk of a 10\,m telescope at the corresponding wavelength.

\bibliographystyle{aa}
\bibliography{ArXiv}

\end{document}